\documentclass[twocolumn,oldversion]{aa}
\usepackage{txfonts}

\usepackage{graphicx}
\usepackage{psfig}
\usepackage{natbib}
\bibpunct{(}{)}{;}{a}{}{,}

\newcommand{\QUINT}{{ \rm Q}} 
\newcommand{\START}{{\rm i}}
\newcommand{\END}{{\infty}} 
\newcommand{\MID}{{\rm av}}\newcommand{\TRANS}{{\rm t}}

\newcommand{\UUNIT}[2]{\;{\mbox{#1}}^{#2}}

\newcommand{\MAT}{{\rm m}}

\newcommand{\XXX}{1 + z'(z)}

\begin{document}

\title{What can { be} learned about dark energy evolution?}
\titlerunning{}

\author{Marian Douspis \inst{1} \and
        Yves Zolnierowski \inst{2} \and
        Alain Blanchard \inst{3} \and
        Alain Riazuelo \inst{4} }

\offprints{Marian Douspis}
 
\institute{ Institut d'Astrophysique Spatiale (IAS), Univ. Paris--Sud, CNRS, B\^atiment 121, F-91405 Orsay,  France \\
  \email{marian.douspis@ias.u-psud.fr} \and Laboratoire
  d'Annecy-le-Vieux de Physique des Particules, UMR5814 CNRS, 9 Chemin
  de Bellevue, BP 110,
  F--74941 Annecy-le-Vieux c\'edex, France\\
  \email{zolnierowski@lapp.in2p3.fr} \and LATT , Universit\'e de
  Toulouse, CNRS, 14 avenue \'E.~Belin, F--31400 Toulouse, France \\
  \email{alain.blanchard@ast.obs-mip.fr} \and CNRS, UMR 7095, Institut
  d'Astrophysique de Paris, Paris, F-75014 France ; Universit\'e
  Pierre et Marie Curie-Paris 6, UMR 7095, Paris, F-75014 France \\
  \email{riazuelo@iap.fr} }

\date{10/2007}

\abstract{ We examine constraints obtained from SNIa surveys on a two
  parameter model of dark energy in which the equation of state $w (z)
  = P(z) / \rho (z)$ undergoes a transition over a period
  significantly shorter than the Hubble time. We find that a
  transition between $w \sim -0.2$ and $w \sim -1$ (the first value
  being somewhat arbitrary) is allowed at redshifts as low as $0.1$,
  despite the fact that data extend beyond $z \sim 1$.  Surveys with
  the precision anticipated for space experiments should allow only
  slight improvement on this constraint, as a transition occurring at
  a redshift as low as $\sim 0.17$ could still remain
  undistinguishable from a standard cosmological constant.  The
  addition of a prior on the matter density $\Omega_\MAT = 0.3$ only
  modestly improves the constraints.  Even deep space experiments
  would still fail to identify a rapid transition at a redshift above
  $0.5$.  These results illustrate that a Hubble diagram of distant
  SNIa alone will not reveal the actual nature of dark energy at a
  redshift above $0.2$ and that only the local dynamics of the
  quintessence field can be infered from a SNIa Hubble
  diagram. Combinations, however, seem to be very efficient: we found
  that the combination of present day CMB data and SNIa already excludes
  a transition at redshifts below $0.8$.

\keywords{Cosmology -- Cosmic microwave background -- Supernovae --
Cosmological parameters}}

\maketitle

\section{Introduction}

The nature of dark energy is one of the most puzzling
mysteries of modern cosmology. It is now widely accepted that our
universe is experiencing a phase of accelerated
expansion~\citep{quintinf}. The evidence was first found using a type Ia
supernovae luminosity vs.\ redshift diagram~\citep{riess98,perl98},
but a number of other observations now support this conclusion. In
particular, estimations of the matter density of the universe
generally lead to a low value, while the CMB anisotropies point toward
a spatially flat Universe~\citep{Lineweaver1997,melnat2000}.  WMAP
data also require the presence of dark energy~\citep{wmap2par} unless
one considers an unexpectedly low value of the Hubble parameter $H_0
\leq
50\UUNIT{km}{}\UUNIT{s}{-1}\UUNIT{Mpc}{-1}$~\citep{bl03,HuntSarkar07}.
The possible correlation of the CMB fluctuation map with surveys of
extragalactic objects~\citep{corrisw,corrisw2,corrisw3} also provides
direct evidence, although with a limited significance level ($<
3\sigma$), for the existence of an unclustered dark energy component
in the universe, the detected correlation being explainable through
the integrated Sachs-Wolfe effect, i.e., a time variation of the
gravitational potential, which is achieved only if the (baryonic +
cold) matter density parameter $\Omega_\MAT $ { significantly
  differs from 1}.  Finally, the shape of the correlation function on
scales up to $100 h^{-1}$ Mpc which has been recently measured
accurately~\citep{eisenstein2005} in combination with the CMB data
{ advocates} for the presence of dark energy~\citep{bl05} { in
  the framework of general relativity}.  However, the nature of this
dark energy has been the subject of numerous speculations. The
simplest model, which was originally proposed (in another
context) by~\citet{eins17}, is a pure cosmological constant $\Lambda$,
a term on the left hand side of Einstein's equations. However, a
cosmological constant can also be regarded as the contribution of the
vacuum to the right hand side of the equation with a specific equation
of state, i.e., a component with negative pressure $P_\Lambda$ related
to the energy density $\rho_\Lambda$ by the relation $P_\Lambda = -
\rho_\Lambda$.  Indeed, quantum field theory predicts that the lowest
energy state of any mode contributes to a vacuum energy density that
behaves exactly as a cosmological constant~\citep[see, e.g.][]{vac}. A
number of problems arise with this possibility, in particular the
so-called hierarchy problem: the expected contribution is usually
enormous, naive calculation gives $\rho_{\rm vac} \sim
10^{76}\UUNIT{GeV}{4}$, around 122 orders of magnitude larger than the
present critical density of the universe. However, there exist 
mechanisms, such as supersymmetry, which allow one to reduce considerably
the vacuum energy density, but since supersymmetry is broken at a
scale larger than $100\UUNIT{GeV}{}$ one is still plagued with an
enormous vacuum energy density of the order of $10^8\UUNIT{GeV}{4}$. The
usual explanation is then to say that there exists a yet unknown
mechanism which ensures that the contribution of the vacuum energy
density is zero. One is therefore left 
to explain the nature of dark energy, which differs from a
cosmological constant, avoiding the extreme fine tuning
required to obtain the observed dark energy density~\citep{vac}. Most
models that have been proposed so far (quintessence models) therefore rely
 on the idea that some scalar field~\citep{quint} behave
today like a cosmological constant, exactly as an other scalar
field did during inflation. 
The most remarkable feature of
quintessence models is that both the scalar field pressure $P_\QUINT$
and energy density $\rho_\QUINT$ evolve according to dynamical
equations. Consequently, the so-called equation of state parameter,
$w_\QUINT \equiv P_\QUINT / \rho_\QUINT$, varies with time between $1$
and $-1$, as the field evolves along the potential.  In some extreme
(and possibly ill-defined) models, this parameter can even take any
arbitrary value, for example if one allows the density $\rho_\QUINT$
to take negative values or a change in the sign of the kinetic
term~\citep{phantom}. Other models involving scalar tensor theories
also allow for such  transient behaviour~\citep{st1}. { The
  detection of such a variation would therefore be of great
  importance for our understanding of dark energy.}

{ The aim of the present paper is to study models with a rapid
  transition of the equation of state and to
illustrate that in this case, the Hubble diagram of SNIa provides
surprisingly weak constraints compared to the case of a smooth
transition.
In Sec.~\ref{secpara}, we recall
a few basic aspects of simple quintessence models, and the motivation for  a 
convenient parametrization of the equation of state parameter $w_\QUINT (z)$
allowing rapid transition. In Sec.~\ref{secanalysis}, we describe the
 analysis we perform, and  state our main
results.  In Sec.~\ref{secliv} we discuss the crucial issue of 
 the impact of the epoch of observation on the parameter
estimation.  We draw the main conclusions of our work in
Sec.~\ref{secconc}.}

\section{Dark energy parametrization}
\label{secpara}

Historical quintessence models rely on the idea of a tracking
solution~\citep{quint888,quint88}, which involves a scalar field
evolving in an inverse power law potential, $V(Q) \propto
Q^{-\alpha}$, the proportionality constant being tuned so as to obtain
the desired value of the dark energy density parameter $\Omega_\QUINT
\sim 0.7$ today. The main feature of these models is that the pressure
to energy density ratio, $w_\QUINT$, remains constant both in the
radiation era and in the matter era (with different values during each
epoch), and that it tends toward $-1$ once the quintessence energy
density dominates. The value of the parameter $w_\QUINT$ depends on
the power law index of its potential.  From existing data, it seems
that only values close to $w_\QUINT \sim -1$ today, or even possibly
lower, are acceptable~\citep{phantom,melch}. Note however that in the
latter case, values of $w_\QUINT$ below $-1$ cannot be obtained
naturally through a standard scalar field.  Single power law
potentials suffer from the fact that, once quintessence dominates, the
$w_\QUINT$ parameter approaches the asymptotic value $-1$ very slowly
so that today, if the quintessence density parameter $\Omega_\QUINT$
is close to $0.7$, then $w_\QUINT$ is still far from the value $-1$,
contrary to what most analyses suggest.  In order to avoid this
problem, one has to add extra features in the potential, such as a
rapid change in the slope of the potential or a local minimum, such as
in the SUGRA model proposed by~\citet{braxmartin99}. Many other
possibilities have been proposed since then~\citep[see for example
  references in][]{refq,quintinf}.

On the other hand, without precise ideas about the correct
quintessence model, it has become natural to adopt a more
phenomenological approach in which one parametrizes the functional
form of $w_\QUINT(z)$ which exhibits the main features
described above. 

The simplest model of quintessence (in the sense that it introduces
only one new parameter as compared to a $\Lambda$CDM model) is to
assume a constant $w_\QUINT$. However there is little motivation for
constant $w_\QUINT$ beyond the economical argument and it is
increasingly recognized that evolving $w_\QUINT$ should be
investigated with a minimal number of priors.  In the absence of well
motivated theoretical considerations one is left with the empirical
option to examine constraints on the analytical form for $w_\QUINT(z)$.
Most investigations have been based on expressions with one or two
parameters.  However,  such expressions
often vary with time in a relatively slow way and that rapidly varying
expressions have to be examined as well.  In other words, if one
considers the typical time scale:
\begin{equation} 
\tau_\QUINT \sim \frac{w}{\dot{w}} ,
\end{equation} 
constant $w$ corresponds to $\tau_\QUINT \gg t_H$ where $t_H = 1 / H$
is the Hubble time, a smoothly varying expression such as the inverse
power law potential corresponds to $\tau_\QUINT \sim t_H$ and a more
rapidly varying $w$ correspond to $\tau_\QUINT \ll t_H$, such as in the
SUGRA model. Our aim is primarily to investigate constraints on models
for which $\tau_\QUINT \ll t_H$.  This lead us to use the
following model which allows arbitrary rapid transitions and in
which the dark energy $w_\QUINT$ parameter evolves as a function of
the scale factor $a$ according to
\begin{equation}
\label{wqz}
w_\QUINT(a) = \frac{1}{2} (w_\START + w_\END) -
\frac{1}{2} (w_\START - w_\END) \tanh \left(\Gamma
\log\left(\frac{a}{a_\TRANS}\right) \right) .
\end{equation}
The $w$ parameter goes from $w_\START$ at early times to $w_\END$ at
late times, the transition occurring at $a_\TRANS$. 
The transition occurs at redshift $z_\TRANS = 1 /
a_\TRANS - 1$ (a negative value of which corresponds to a transition
in the future) and lasts of
the order of $\Gamma^{-1}$ Hubble times; $\Gamma$ is therefore a parameter
describing the speed of the transition : high values ($\gg  1$) correspond 
to fast transitions, in the limit  $\Gamma = \infty$ the transition is 
instantaneous.  This expression was proposed by \citet{pca4}.
The quintessence conservation equation
$D_\mu T_\QUINT^{\mu\nu} = 0$ can be integrated exactly to give
\begin{equation}
\label{rhoQ}
\rho_\QUINT(a)
 = \rho_\QUINT^0 
   \left(\frac{a}{1}\right)^{-3(1+w_\MID)}
   \left(
         \frac{  \left(a/a_\TRANS\right)^\Gamma
               + \left(a/a_\TRANS\right)^{-\Gamma} }
              {  \left(1/a_\TRANS\right)^\Gamma
               + \left(1/a_\TRANS\right)^{-\Gamma}}
   \right)^{\frac{3 \Delta w}{2 \Gamma}} ,
\end{equation}
where we have set
\begin{equation}
w_\MID \equiv \frac{1}{2} (w_\START + w_\END) , \qquad
\Delta w \equiv (w_\START - w_\END) ,
\end{equation}
and where $\rho_\QUINT^0$ corresponds to the values of the
quintessence energy density at the present epoch. Therefore, the model
can be implemented without much modification in existing cosmological
codes such as CAMB~\citep{CAMB}.  Dark Energy perturbations are
implemented accordingly to \cite{DEpert, DEpert2}, i.e. the initial
spectrum of the dark energy perturbation follows an attractor
mechanism, exactly as the unperturbed part of the DE fluid does, and
the DE initial spectrum is lost. One modification concerns the sound
speed defined as $\dot P / \dot \rho$ which is equal to the $w$
parameter when $w$ is constant. When $w$ is not constant, as is the
case for our model, its exact value is known since the analytic forms
of $w$ and $\rho$ are known. This model has four parameters. As we
said earlier, simple quintessence model have fewer: constant $w$
models correspond to $w_\START = w_\END$, with undefined $a_\TRANS$
and $\Gamma$. With an inverse power law potential, one has $w_\END =
-1$, $w_\START = - 2 / (\alpha + 2)$, the epoch of transition is fixed
approximately by the constraint $\Omega_\QUINT(z_\TRANS) \sim
\Omega_\MAT(z_\TRANS) \sim 0.5$ and the duration of the transition is
larger than the Hubble time (it depends on how steep the potential is,
that is, on $\alpha$). For the SUGRA potential the first two above
constraints on $w_\START$ and $w_\END$ remain, whereas the latter are
modified: the epoch of transition for $w_\QUINT$ no longer necessarily
corresponds to the scalar field domination (but corresponds to the
epoch where the field reaches a local minimum in its potential), and
the transition duration is usually much shorter.  Various other models
have different predictions concerning these parameters, but the above
parametrization is sufficiently general to encompass a large number of
already proposed models.

One of these parameters, $w_\START$, is not expected
to be as relevant as the parameter $a_\TRANS$: because at early times
$\Omega_\QUINT (z)$ is negligible  compared to $\Omega_\MAT$,
the quintessence field does not play a crucial role, at least with respect
to supernovae and CMB data.
One should impose a value of $w_\START$
slightly lower than 0, in order to ensure that at early times
$\Omega_\QUINT \ll \Omega_\MAT$.  The reason is that with $w_\START =
0$ and a low $z$ transition $\Omega_\QUINT$ would be close to its
present value at the recombination epoch~\citep[or even nucleosynthesis,
see references in][]{quintinf}.  At low redshift, this would lead to a
dramatic suppression of the cosmological perturbation growth
rate~\citep{nous2003}. In addition, we found that this introduces
additional changes in the $C_l$ curve at high $l$ (i.e., other than
changes due to the modification of the angular distance).  For these
reasons we fix  $w_\START = -0.2$.  Putting a
constraint on $w_\END$ is less desireable since it implicitly
selects a limited class of models, which do not seem excluded by the
data. We have chosen the value $w_\END = -1$, which
seems in agreement with the present data, and we focus on the two
remaining parameters, $\Gamma$ and $a_\TRANS$ which describe the
transition experienced by $w_\QUINT (z)$ between its early and late
behaviour. 

\section{Analysis}
\label{secanalysis}

We focus here on constraints that can be set in the transition
parameters $z_\TRANS$ and $\Gamma$, and we set $w_\START = -
0.2 $ and $w_\END = -1$ as explained above.  Note that a pure
cosmological constant behaviour is obtained by considering large
$z_\TRANS$ with a sufficiently small transition duration (so that it
does not last long after $z_\TRANS$).

\subsection{Supernovae Hubble diagram}

{ The luminosity distance is one of the main  sources of constraint
on the nature of dark energy (\cite{Astier01}).  We therefore } first
examine what kind of constraints the Supernovae Hubble diagram allows.
The number of well observed SNIa has rapidly increased in recent
years and a significant number of supernovae above redshift one have
been detected.  { In the following we use the last compilation
from~\cite{Davis07} of recent SNIa (\cite{Wood07,Riess07,Astier06}).}

\begin{figure}[!t]
\centerline{\psfig{file=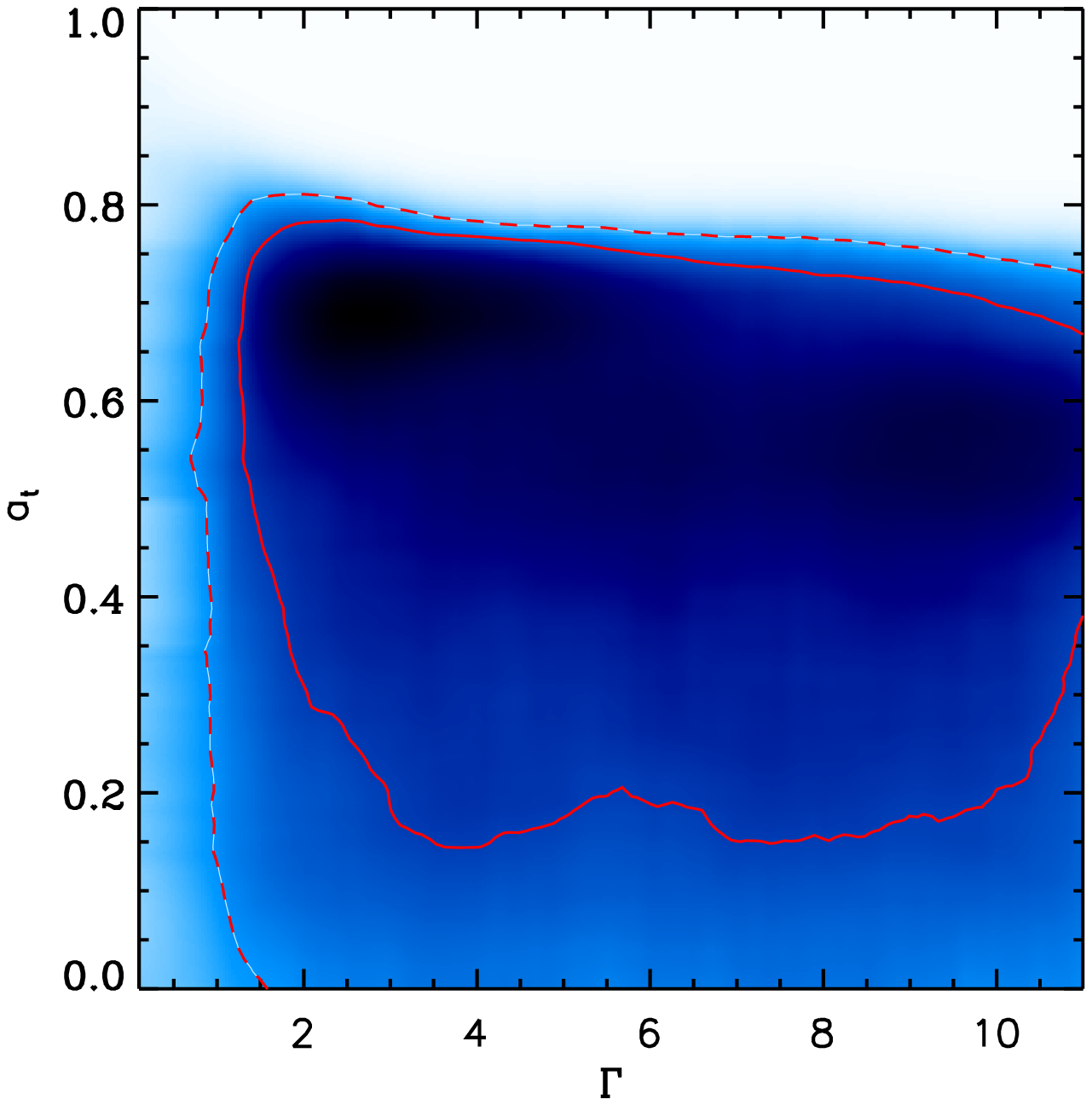,width=3.5in}}
\centerline{\psfig{file=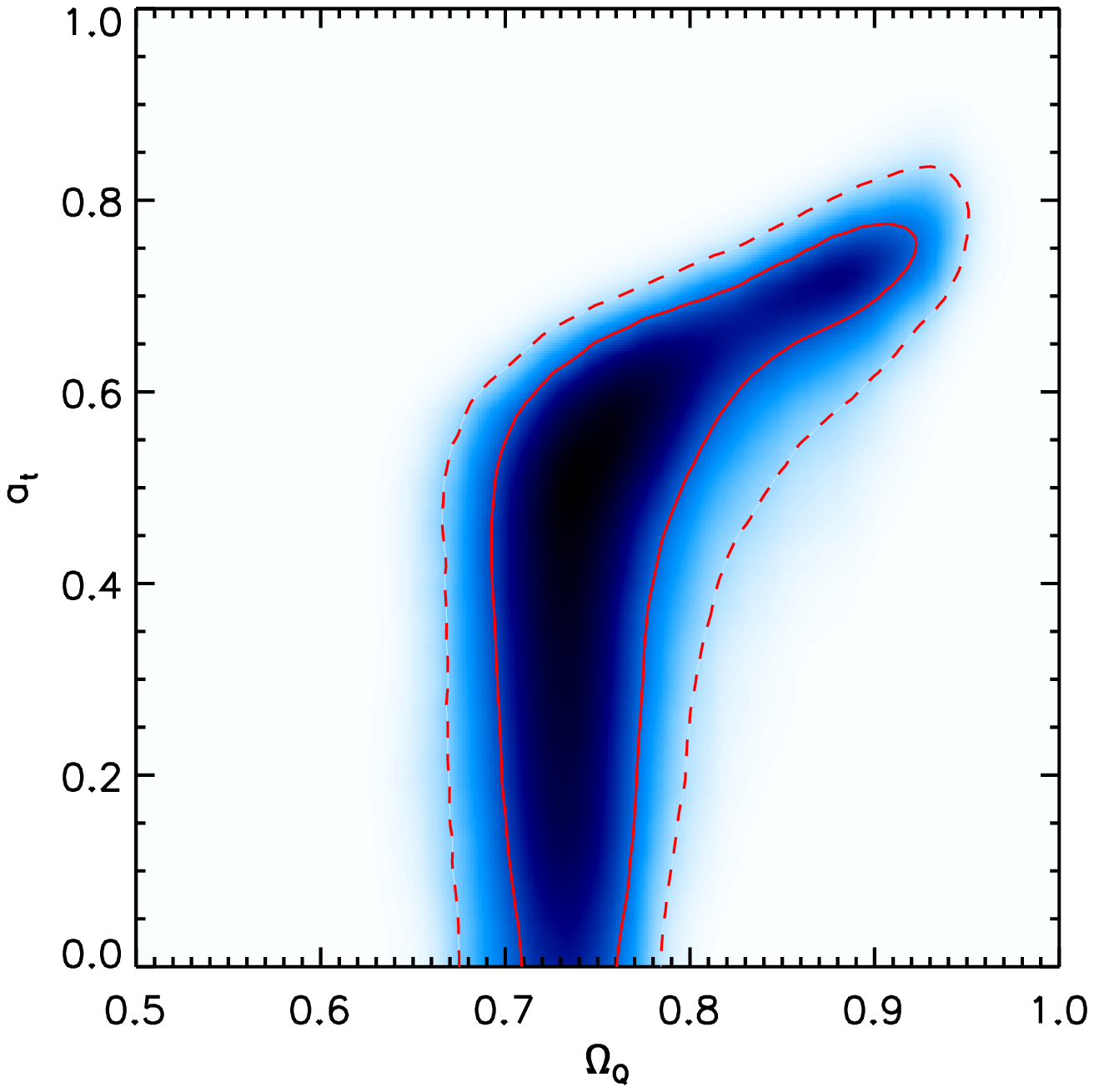,width=3.5in}}
\caption{Contour constraints on the transition at epoch $a_\TRANS = 1
/ (1 + z_\TRANS)$, and the rate of the transition $\Gamma$.
 The
results are independent of $\Gamma$ when it is sufficiently
large.
Bottom figure shows the results
in the ($\Omega_\QUINT$, $a_\TRANS$) plane.}
\label{fig:CD1}
\end{figure}

We have examined constraints on our three parameters, $a_\TRANS = 1 /
(1 + z_\TRANS)$, $\Gamma$ and $\Omega_\QUINT$. { We  use the
publicly available Monte Carlo Markov Chains code, cosmomc (\cite{cosmomc}),
using the modified version of CAMB described above. 2-D contours shown
in Figures~\ref{fig:CD1} and~\ref{fig:CD2} encompass 68\% and 95\% confidence
levels (CL). In each figure, the third parameter is marginalised over.}

The upper graph of Figure~\ref{fig:CD1} presents the allowed regions
in the plane: $a_\TRANS = 1 / (1 + z_\TRANS)$ versus the inverse
duration of the transition $\Gamma$. The lower graph gives the
constraints in the $a_\TRANS$--$\Omega_\QUINT$ plane. 
Constraints on possible transitions appear very weak: only sharp
transitions at very low redshift ($z_\TRANS < 0.1$ at the two sigma
level) are firmly excluded.  Surprisingly, the data suggest a
transition at low redshift, a tendency that has been noticed
elsewhere~\citep{basset,kunz2004}. However the significance
level is low and a cosmological constant remains consistent with the
data at the { 2 sigma level}.

While rapid transitions (corresponding to large $\Gamma$) are very weakly
constrained, better constraints are obtained when a strong
prior is set on $\Omega_\MAT$: with $\Omega_\MAT = 0.3$ we found that
transitions are acceptable at redshifts greater than $0.25$.
This improvement is due to the { removing of degeneracy breaking (noticed in
  the parameter space of Figure~\ref{fig:CD1})} but remains modest.
   This means that the Hubble diagram of distant SNIa alone is
   insufficient to determine the nature of the dark energy at high
   redshift.

\begin{figure}[!t]
\centerline{\psfig{file=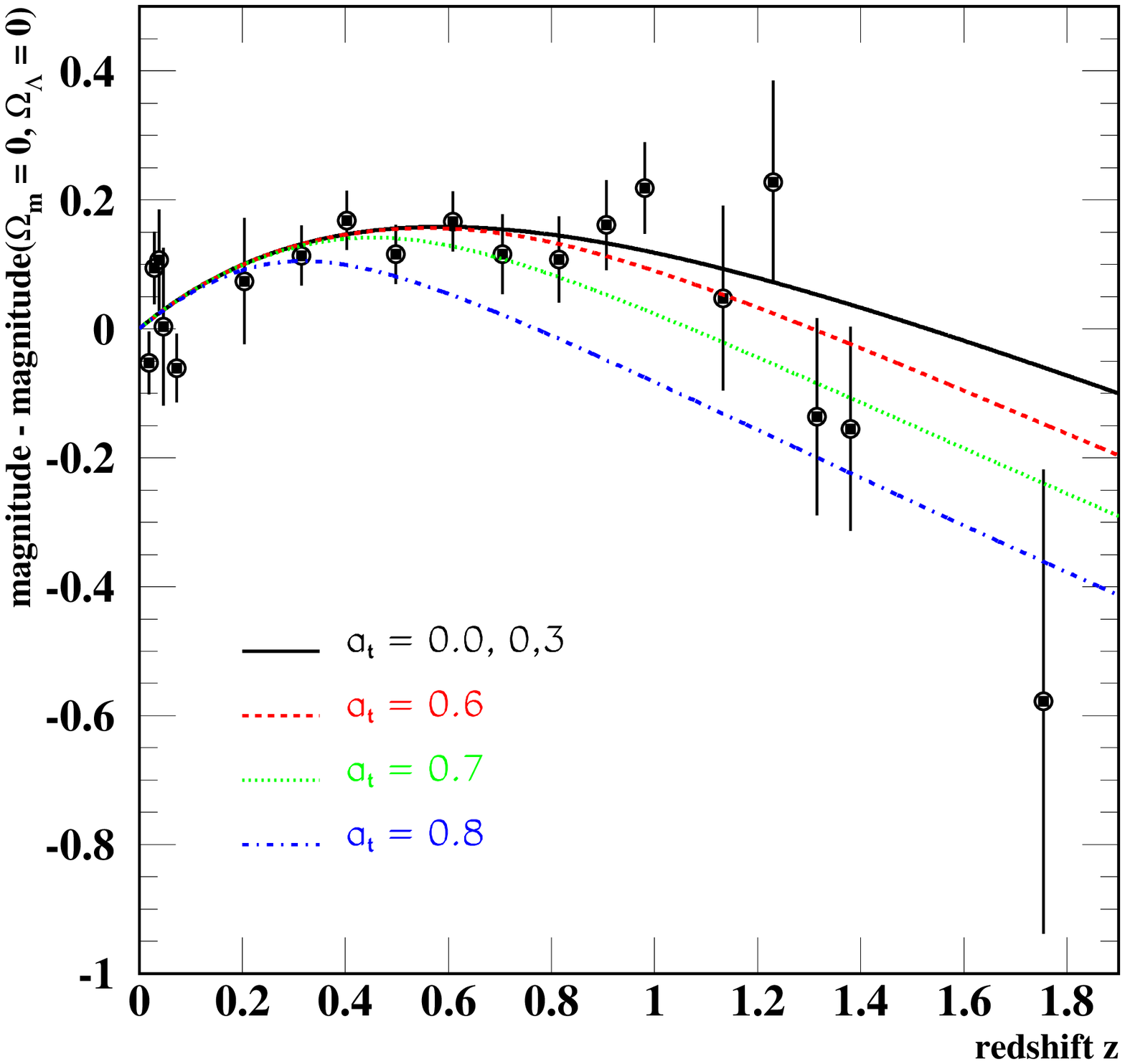,width=8cm}}
\centerline{\psfig{file=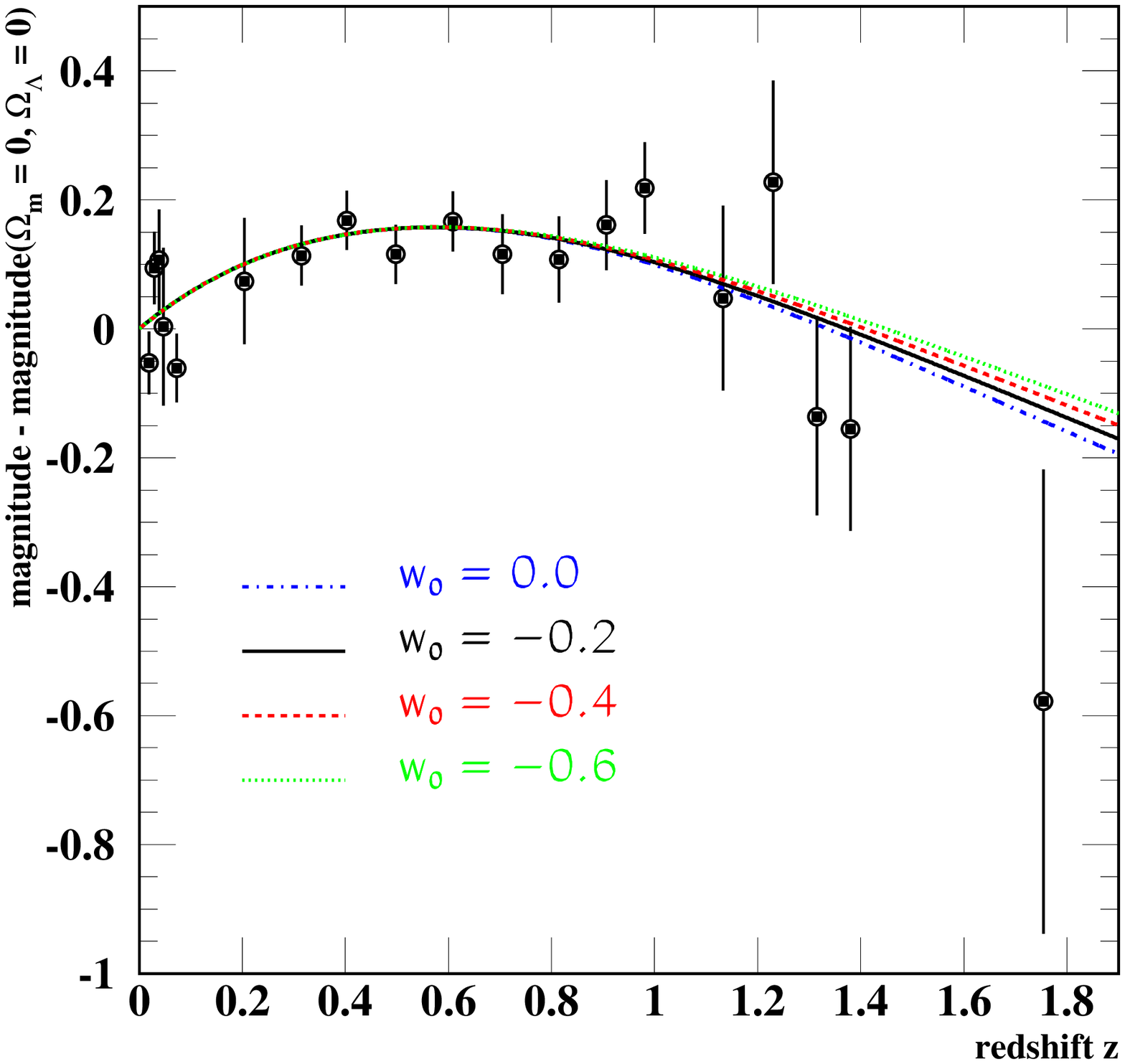,width=8cm}}
\caption{a) Residual Hubble diagram with respect to an empty universe for
  models with a transition at epoch $a_\TRANS = 1/(1+z_\TRANS) =
  0.0$, $0.3$, $0.6$, $0.7$ $\&$ $0.8$ compared with a binned version
  of the SNIa compilation of~\cite{Davis07}. Note that as explained in
  the text low values of $a_\TRANS$ all give similar curves as the
  $\Lambda$CDM model. b) Same quantity for models with a transition at  epoch $a_\TRANS = 
  0.565$ and $w_\START = 0, -0.2, -0.4, -0.6$}
\label{fig:HD}
\end{figure}

In Figures \ref{fig:HD}, we show the effect of a transition on the
magnitude difference between a fiducial model and the empty universe,
{ all other parameters being fixed with their fiducial values}. The
top figure clearly reveals that a transition from $w_\START = -0.2$ to
$w_\END = -1$ occurring at even moderately low redshift makes very
little differences to the observable quantity. It is therefore not
surprising that the constraints that can be set from the present day
SNIa Hubble diagram are not very tight. The bottom figure illustrates
the effect of changing $w_\START$: changing $w_\START$ from $0 $ to
$-0.6$ produces changes that are small and easy to understand as the
model becomes degenerate with the $\Lambda$CDM model as $w_\START$
tends to $-1$. For this reason, in the following, we concentrate our
analysis on $a_\TRANS$.

 We have redone the above analysis on $a_\TRANS$ for a simulated
 survey with the precision and statistics expected from space
 experiments. { We generated 2000 supernovae distributed in 16 bins in
   redshift between 0.2 and 1.7 according to Table 1 of \cite{kim04}
   completed by 300 nearby supernovae. The number of supernovae per
   bin fluctuates according to a Poisson law.  For a given bin, the
   magnitude of the supernovae is taken from a Gaussian distribution
   of the mean value given by the standard concordance $\Lambda$CDM
   model, and the sigma fixed to 0.2 magnitude.  The resulting
   magnitude of each bin is obtained by a fit of the distribution of
   magnitudes and the associated error is added in quadrature with a
   systematic error of 0.02 magnitude and an offset error of 0.01
   magnitude for the intercalibration between the two sets of data.}
The constraints inferred from this simulated sample again reveal  that
the transition epoch $a_\TRANS$ is moderately constrained: transitions
at redshift as low as $0.5$ (2 $\sigma$ CL) are still acceptable when
a rapid transition ($\Gamma > 2$) is assumed.

The situation is therefore  paradoxical: although space survey
precision improves the constraints by pushing the acceptable redshift
{ from 0.25 to 0.5 (for rapid transitions)}, this last result
is modest as a significant fraction of the
high precision data provided by the experiment extends up to redshift
$\sim 2$.  The reason for this apparent paradox is clarified in
Sec.~\ref{secliv}.

\subsection{Constraints from the CMB}

\begin{figure}[!t]
\centerline{\psfig{file=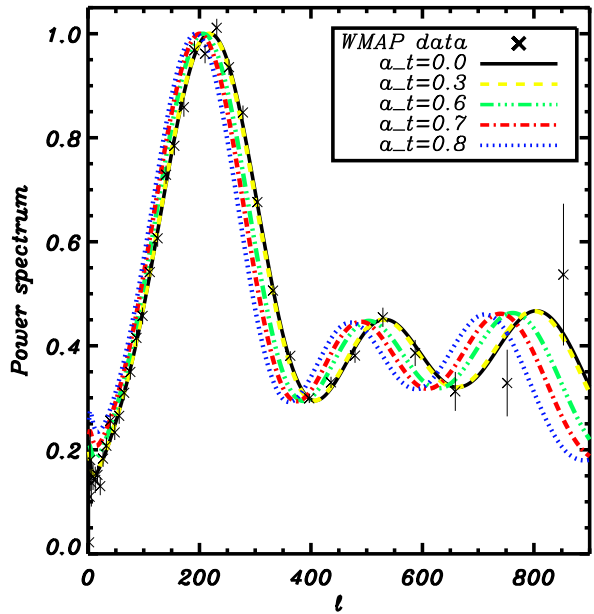,width=3.5in}}
\caption{Angular power spectrum of CMB fluctuations for models
presenting a transition at epoch $a_\TRANS = 1 / (1 + z_\TRANS) = 0$,
$0.3$, $0.6$, $0.7$ $\&$ $0.8$ { (all other parameters fixed)}
compared with the WMAP 3 TT spectrum. Apart from the large scale (low
$\ell$) power, the only difference between the spectra is a shift in
the spectrum as a result of the modification of the angular distance
to the last scattering surface.}
\label{fig:cmbCl}
\end{figure}

Given that SNIa Hubble diagram hardly suggests the presence of 
a transition in the dark energy content of the universe, it is interesting 
to examine whether such a possibility could be acceptable  using 
additional constraints.  
CMB is known to provide such  constraints on the quintessence
scenario, eg~\citep{spergel03,nous2003,odman2004}.
We therefore  examine CMB constraints on the type of
models introduced above, although we leave to a future work a full
investigation of the constraints that can be set on this type of
model. We use the WMAP 3 data-set, as well as
CBI, VSA and  Boomerang data at small scales, and a version of the CAMB
cosmological code~\citep{CAMB} that we have modified. Modifications of
the code are  straightforward since its public version includes
models with constant $w_\QUINT$ in which we have implemented
the energy density
$\rho_\QUINT (z)$ and the pressure $P_\QUINT(z)$ as a function of
redshift. Our ansatz for the equation of state parameter
$w_\QUINT(z)$ allows us to integrate the conservation equation to obtain
an analytical form for $\rho_\QUINT(z)$\footnote{One has
also to make the distinction between the equation of state
parameter $w_\QUINT = P_\QUINT / \rho_\QUINT$ and the ``sound speed''
squared $c_{s\,\QUINT}^2 \equiv \dot P_\QUINT / \dot \rho_\QUINT$,
which are identical when $w$ is constant.}.

\begin{figure}[!t]
\centerline{\psfig{file=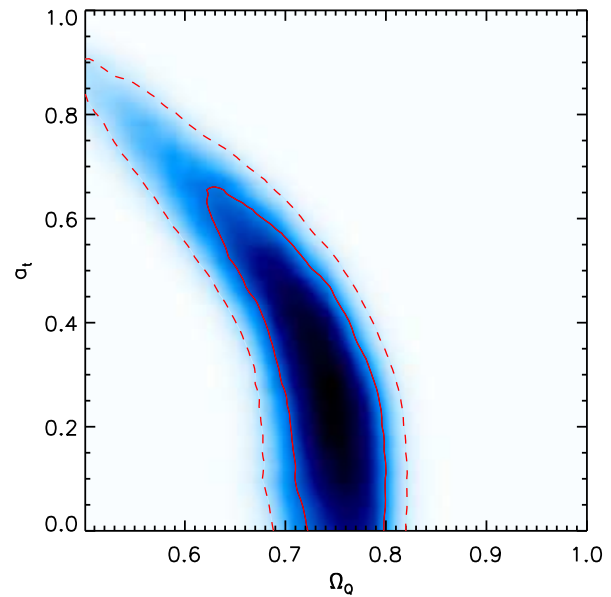,width=3.5in}}
\centerline{\psfig{file=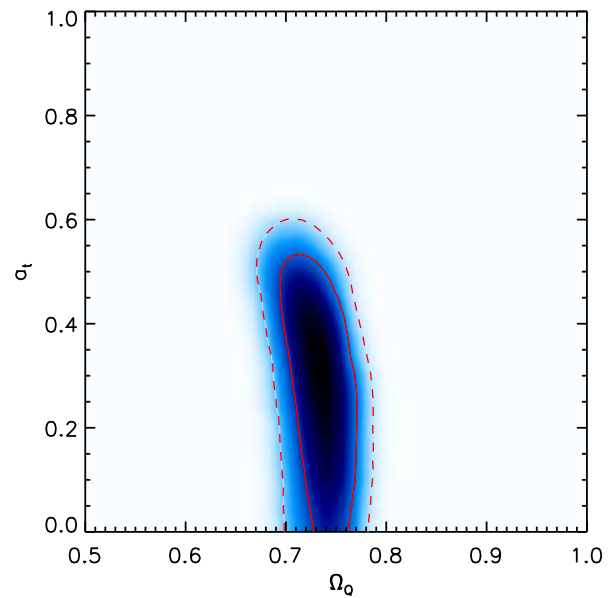,width=3.5in}}
\caption{Contour constraints on the transition epoch $a_\TRANS / a_0 =
1 / (1 + z_\TRANS)$ versus the value of the dark energy density
parameter at present time, $\Omega_\QUINT$ from the CMB alone (upper) using
 WMAP3 , CBI, VSA and Boomerang data. Constraints on the same quantities when
combined with supernovae data (lower).}
\label{fig:CD2}
\end{figure}

The angular power spectrum of CMB fluctuations in the presence of dark
energy is modified mainly through the modification of the angular
distance~\citep{bl84} (see Fig.~\ref{fig:cmbCl} { and \cite{elga07}}).
Although a strong dependence appears, this is partially lost through
parameter degeneracies which strongly weaken the final constraints. In
addition, ISW will contribute to lower levels as the transition is
assumed at lower redshift, and this effect contributes to modify the
angular power spectrum of the CMB fluctuations.  We have investigated
the CMB constraints on models with rapid transitions described by
Eq.~\ref{wqz}.  $\Gamma$ was set to 10.  We have checked that varying
$\Gamma$ above 5 produces no appreciable differences. The other
parameters that were left free are { the baryon budget $\Omega_B$, the
  optical depth $\tau$,} the Hubble constant $H_0$, the dark energy
density at the present day $\Omega_\QUINT$, the index of the
primordial spectrum $n$, the amplitude of fluctuations $\sigma_8$ and
the transition epoch $a_\TRANS = 1 / (1 + z_\TRANS)$. From the
contours obtained in figure \ref{fig:CD2}, one can see that the
constraints that can be set on the transition redshift $z_\TRANS $
from the CMB are rather stringent, $z_\TRANS > 0.54 $ (1$\sigma$ on
one parameter) when $\Omega_\QUINT$ represents less than $50\%$ of the
total density. These constraints being slightly dependent on $H_0$ we
have also examined whether a combination of CMB and supernova data allows
to improve the transition epoch constraints, but although the SNIa
data restricted the dark energy density much more around
$\Omega_\QUINT \sim 0.7$, the final constraints do not represent a
significant improvement: the final constraint {shown in
  Figure~\ref{fig:CD2}} is $z_\TRANS > 0.66 $ (2$\sigma$ on one
parameter). Some different dark energy
  models could in principle lead to different conclusions in the case
  where the sound speed varies in a way that significantly affects the
  integrated Sachs-Wolfe effect on large angular scales, even though
  no such model was found in our analysis. Clearly, better
constraints could be obtained from additional data of cosmological
relevance, but this is beyond the scope of the present paper.

\section{Observing the Universe at $z= 0$ and at $z = 0.3$}
\label{secliv}

In order to distinguish quintessence models from a pure cosmological
constant, it is crucial to be able to track the dark energy evolution
as early as possible.
The main
impact of dark energy comes from its influence on the expansion rate
of the universe.
An important question is therefore until what epoch the dark
energy density  plays a role in observable quantities, and as a
corollary, until what epoch one can hope to reconstruct either its
energy density or its equation of state parameter. As we have seen in
section \ref{secanalysis}, the SNIa Hubble diagram poorly constrains a
possible transition epoch in the equation of state of the dark energy
component. As we have stated, this appears somewhat paradoxical as
data extending up to redshift 2 fail to reveal a transition occurring
at redshifts as low as 0.25, at which the dark energy component is still
dominant. Indeed, in a model with $\Omega_\QUINT = 0.7$, $\Omega_\MAT
= 0.3$ today, with $w_\QUINT$  constant and equal to $-1$,
the matter to dark energy transition, defined when $\Omega_\QUINT (z)
= \Omega_\MAT(z) = 0.5$, occurs at redshift $z_e = (\Omega_\QUINT /
\Omega_\MAT)^{\frac{1}{3}} - 1 \sim 0.33$. Let us now consider two
alternatives. First, we can consider a pure cosmological constant
model, with $w_\QUINT = -1$ also at early times. Second, we can
consider a model where $w_\QUINT \sim 0$ for $z > z_\TRANS = z_e$. In
the first case, one has a usual $\Lambda$CDM model, whereas in the
second case, one has a model close to a flat Einstein-de~Sitter model
at epoch $z > z_e$. An observer  at $z = z_\TRANS$
should easily be able to distinguish between the two models, just as
we are able to distinguish between a $\Lambda$CDM with $\Omega_\QUINT
= 0.5$ and a flat Einstein-de~Sitter model today. Now, are we able to
distinguish today between these two models, which differ only in $z >
z_\TRANS$?  Surprisingly, the answer is no if one considers supernovae
data only, as  is convincingly illustrated by Fig.\ref{fig:HD}. The
explanation of this apparent paradox is as follows. Present data
favour dark energy because high redshift supernovae are dimmer than
expected in a flat Einstein-de~Sitter universe. This is usually
expressed as a difference of magnitude between the two models one
considers for some standard candle at some redshift, the exact value
of which depend on the quality of the data. The magnitude is
essentially the logarithm of the luminosity distance as a function of
the redshift. Let us define $d_L^\Lambda(z)$ and $d_L^{\rm EdS}(z)$
the luminosity distance as a function of the redshift in a
$\Lambda$CDM model with $\Omega_\Lambda = \Omega_\MAT = 0.5$ today,
and in a flat Einstein-de~Sitter model. Let us assume these
two models can be distinguished. Let us now consider $\tilde
d_L^\Lambda(z)$ and $d_Q(z)$ the luminosity distance vs.\ redshift
relation in a $\Lambda$CDM model with $\Omega_\Lambda = 0.7$,
$\Omega_\MAT = 0.3$ today, and a dark energy model with $\Omega_Q =
0.7$, $\Omega_\MAT = 0.3$ today, with $w_\QUINT$ experiencing a sudden
transition from $0$ to $-1$ at $z = z_\TRANS$. An observer  at
$z = z_\TRANS$ would therefore measure either $d_L^\Lambda(z')$ or
$d_L^{\rm EdS} (z')$. The epoch corresponding to a redshift of $z'$
measured by an observer  at $z_\TRANS$ corresponds to a redshift
$z$ given by
\begin{equation}
z = (1 + z_\TRANS) (1 + z') - 1 = z_\TRANS + z'+ z_\TRANS z',
\end{equation}
measured by an observer  today. Let us define $d_\TRANS$ as the
luminosity distance of the observer  at $z = z_\TRANS$ as seen
from today. The exact value of $d_\TRANS$ does not matter here, but it
can be  computed as
\begin{equation}
d_\TRANS = - d_L^\Lambda \left(\frac{1}{1 + z_\TRANS} - 1\right) .
\end{equation}
Luminosity distances do not add but are proportional to comoving
distances.  The luminosity distance at redshifts above $z_\TRANS$ is
therefore given by
\begin{eqnarray}
\tilde d_L^\Lambda (z) & = &
  \frac{1 + z}{1 + z_\TRANS} d_\TRANS
 +\frac{1 + z}{\XXX} d_L^\Lambda(z'(z)) , \\
d_L^Q (z) & = &
  \frac{1 + z}{1 + z_\TRANS} d_\TRANS
 +\frac{1 + z}{\XXX} d_L^{\rm EdS}(z'(z)) , 
\end{eqnarray}
in the two models. Now, in term of difference of
magnitude between the two models, this is translated into
\begin{equation}
\label{dmt}
\Delta \tilde m (z) \propto 
\log \frac{d_t + \frac{1 + z_\TRANS}{\XXX} d_L^\Lambda(z'(z))}
          {d_t + \frac{1 + z_\TRANS}{\XXX} d_L^{\rm EdS}(z'(z))} ,
\end{equation}
whereas for the observer  at $z_t$, this simply gives
\begin{equation}
\label{dm}
\Delta m (z') \propto 
\log \frac{d_L^\Lambda(z')}{d_L^{\rm EdS}(z')} .
\end{equation}
Two differences arise here. First, the redshift range which today's
observer must use in order to distinguish between the two models is
larger than that of the observer  at $z_\TRANS$. If the latter
must collect data between $z'= 0$ and $z'= z_*$, say, then the former
must observe supernovae between $z(0) = z_\TRANS$ and $z(z_*) =
z_\TRANS + z_* + z_\TRANS z_* > z_\TRANS + z_*$. Second, the magnitude
difference between the two models is strongly attenuated for the
observer  today because of the presence of the extra term
$d_\TRANS$ both in the numerator and the denominator of
Eq.~(\ref{dmt}). { This behavior is somewhat surprising at first
   but can be understood from the expression of the comoving
  coordinate $r$:
\begin{equation}
r(z) = \int _0^z\frac{dz}{H(z)}
\end{equation}
where $H(z)$ follow the Fridmann-Lema\^{\i }tre equation:
\begin{equation}
H^2(z) = 8\pi G{\rho_m+\rho_Q}
\end{equation}
(for $k =0$). Let us assume that a rapid transition occurs at $z =
z_\TRANS$ between $w = -1$ and $w \sim 0$; it is straightforward to
compute the relation between the Hubble constant $\tilde{H}(z)$ in
such a model to the Hubble constant $H(z)$ in the standard model:
\begin{equation}
  \frac{\tilde{H}^2(z)}{H^2(z)}= \frac{1+\frac{1.-\Omega_0 }{\Omega_0(1+z_\TRANS)^3} }{1+\frac{1.-\Omega_0 }{\Omega_0(1+z)^3} }.
\end{equation}
It is clear from this expression that the values of the Hubble
constant in the two models do not differ by much when the
transition redshift is not close to zero. As the integrand in the
calculation of $r$ is higher at low redshift, the limited difference
in $H$ translates to a  small  difference in $r$.  For instance a
transition at $a_\TRANS \sim 0.7$ results in a 5\% decrease in $r$ at
redshift 1 corresponding to $\Delta m \sim 0.1$ as can be seen in
figure \ref{fig:HD}.}

\begin{figure}
\centerline{\psfig{file=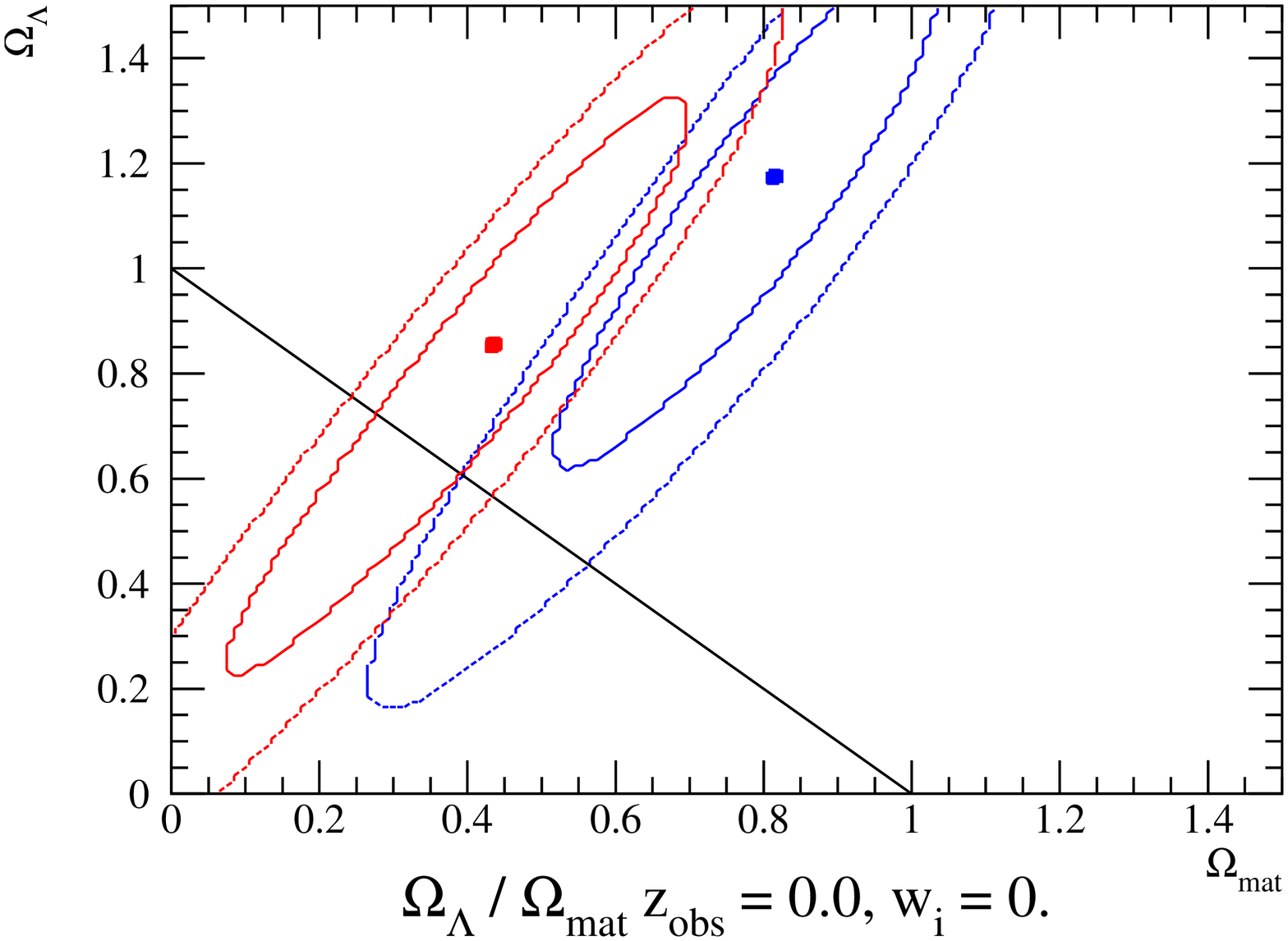,width=8cm}}
\centerline{\psfig{file=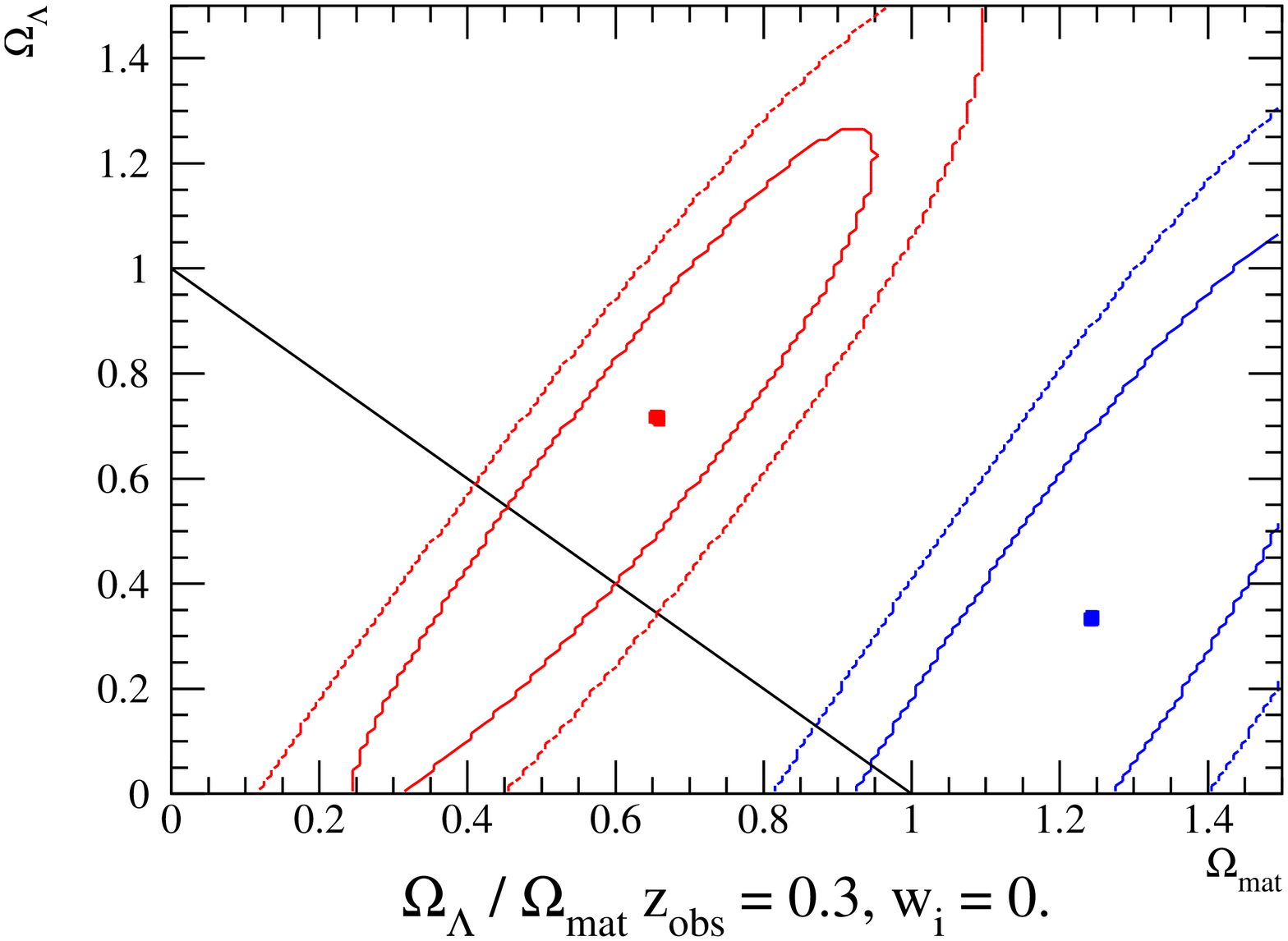,width=8cm}}
\caption{Comparison of $\Omega_\MAT-\Omega_\Lambda$ fit for the two
models discussed in Sec.~\ref{secliv} for an observer at $z =
0$ (after the transition in the quintessence model) and $z = 0.3$
(during the transition). At $z = 0.3$ the two models are very easily
distinguished as one of them corresponds to a pure matter model,
whereas one corresponds to a cosmological constant model with
$\Omega_\Lambda = 0.5$.  On the contrary, at $z = 0$ both models look
like a cosmological constant model, and cannot be distinguished at the
2$\sigma$ level.}
\label{figcomp}
\end{figure}

The net result is that while both models are easy
to distinguish at $z = 0.3$, this is no longer the case at $z = 0$ as
 seen in Fig.~\ref{figcomp}.

\section{Conclusion}
\label{secconc}

We have investigated a class of models that undergo a rapid
transition in the equation of state of their dark energy component. In
order to establish the constraints that can be obtained on the
characteristics of the transition we have focused our study on a class
of models in which dark energy transits rapidly between $w \sim 0$ and
$w \sim -1$.  We found that the duration of the
transition cannot be constrained when it is shorter than the
Hubble time. More surprisingly we found that SNIa Hubble diagram does
not constrain this type of scenario very much, as even with the data
expected from space experiments a transition can still be allowed at
epochs when the dark energy density represents 40\% of the density of
the Universe. This suggests that the SNIa diagram is  poorly sensitive to
dynamics of dark energy at redshifts above $0.5$.  On the contrary, we
found that existing CMB data, in combination with SNIa data, 
provide tight constraints on this type of scenario, allowing rapid
transitions to happen only at redshifts beyond $0.66$, when dark energy
represents less than 10\% of the total density of the Universe.  {
  This illustrates the importance of combinations of various data in
  order to  accurately constrain  the evolution of dark
  energy.}

\begin{acknowledgements}
  The authors acknowledge Programme National de Cosmologie for financial
  support during the preparation on this work. The authors also
  acknowledge N.Aghanim for her comments. M.D. would like to
  acknowledge the French Space agency (CNES) for financial support and
  Laboratoire d'astrophysique de Toulouse Tarbes where most of the work
  was done. Y.Z. would like to acknowledge IN2P3 and INSU for
  financial support and Laboratoire d'astrophysique de Toulouse Tarbes.
      
\end{acknowledgements}

\bibliographystyle{aa} 

\bibliography{final} 

\begin{thebibliography}{41}
\expandafter\ifx\csname natexlab\endcsname\relax\def\natexlab#1{#1}\fi

\bibitem[{{Astier}(2001)}]{Astier01}
{Astier}, P. 2001, Physics Letters B, 500, 8

\bibitem[{{Astier} {et~al.}(2006){Astier}, {Guy}, {Regnault}, {Pain},
  {Aubourg}, {Balam}, {Basa}, {Carlberg}, {Fabbro}, {Fouchez}, {Hook},
  {Howell}, {Lafoux}, {Neill}, {Palanque-Delabrouille}, {Perrett}, {Pritchet},
  {Rich}, {Sullivan}, {Taillet}, {Aldering}, {Antilogus}, {Arsenijevic},
  {Balland}, {Baumont}, {Bronder}, {Courtois}, {Ellis}, {Filiol}, {Gon{\c
  c}alves}, {Goobar}, {Guide}, {Hardin}, {Lusset}, {Lidman}, {McMahon},
  {Mouchet}, {Mourao}, {Perlmutter}, {Ripoche}, {Tao}, \& {Walton}}]{Astier06}
{Astier}, P., {Guy}, J., {Regnault}, N., {et~al.} 2006, \aap, 447, 31

\bibitem[{Bassett {et~al.}(2004)Bassett, Corasaniti, \& Kunz}]{basset}
Bassett, B.~A., Corasaniti, P.~S., \& Kunz, M. 2004, \apj, 617, L1

\bibitem[{Bin\'etruy(2000)}]{vac}
Bin\'etruy, P. 2000, in The primordial universe, ed. P.~Bin\'etruy,
  R.~Schaeffer, J.~Silk, \& F.~David (EDP Sciences, Paris and Springer, Berlin)

\bibitem[{Blanchard(1984)}]{bl84}
Blanchard, A. 1984, \aap, 132, 359

\bibitem[{Blanchard {et~al.}(2003)Blanchard, Douspis, Rowan-Robinson, \&
  Sarkar}]{bl03}
Blanchard, A., Douspis, M., Rowan-Robinson, M., \& Sarkar, S. 2003, \aap, 412,
  35

\bibitem[{{Blanchard} {et~al.}(2006){Blanchard}, {Douspis}, {Rowan-Robinson},
  \& {Sarkar}}]{bl05}
{Blanchard}, A., {Douspis}, M., {Rowan-Robinson}, M., \& {Sarkar}, S. 2006,
  \aap, 449, 925

\bibitem[{Brax \& Martin(1999)}]{braxmartin99}
Brax, P. \& Martin, J. 1999, Phys.\ Lett.\, 468B, 40

\bibitem[{Brax {et~al.}(2000)Brax, Martin, \& Riazuelo}]{refq}
Brax, P., Martin, J., \& Riazuelo, A. 2000, \prd, 62, 103505

\bibitem[{{Brax} {et~al.}(2000){Brax}, {Martin}, \& {Riazuelo}}]{DEpert2}
{Brax}, P., {Martin}, J., \& {Riazuelo}, A. 2000, \prd, 62, 103505

\bibitem[{Caldwell(2002)}]{phantom}
Caldwell, R.~R. 2002, Phys.\ Lett.\, 545B, 23

\bibitem[{Caldwell {et~al.}(1998)Caldwell, Dave, \& Steinhardt}]{quint}
Caldwell, R.~R., Dave, R., \& Steinhardt, P.~J. 1998, \prl, 80, 1582

\bibitem[{Corasaniti {et~al.}(2005)Corasaniti, Giannantonio, \&
  Melchiorri}]{corrisw2}
Corasaniti, P.~S., Giannantonio, T., \& Melchiorri, A. 2005, \prd, 71, 123521

\bibitem[{Corasaniti {et~al.}(2004)Corasaniti, Kunz, Parkinson, Copeland, \&
  Bassett}]{kunz2004}
Corasaniti, P.~S., Kunz, M., Parkinson, D., Copeland, E.~J., \& Bassett, B.~A.
  2004, Phys.\ Rev.\ D, 70, 083006

\bibitem[{{Davis} {et~al.}(2007){Davis}, {M{\"o}rtsell}, {Sollerman}, {Becker},
  {Blondin}, {Challis}, {Clocchiatti}, {Filippenko}, {Foley}, {Garnavich},
  {Jha}, {Krisciunas}, {Kirshner}, {Leibundgut}, {Li}, {Matheson}, {Miknaitis},
  {Pignata}, {Rest}, {Riess}, {Schmidt}, {Smith}, {Spyromilio}, {Stubbs},
  {Suntzeff}, {Tonry}, {Wood-Vasey}, \& {Zenteno}}]{Davis07}
{Davis}, T.~M., {M{\"o}rtsell}, E., {Sollerman}, J., {et~al.} 2007, \apj, 666,
  716

\bibitem[{de~Bernardis {et~al.}(2000)de~Bernardis, Ade, Bock, Bond, Borrill,
  Boscaleri, Coble, Crill, Gasperis, Farese, Ferreira, Ganga, Giacometti,
  Hivon, Hristov, Iacoangeli, Jaffe, Lange, Martinis, Masi, Mason, Mauskopf,
  Melchiorri, Miglio, Montroy, Netterfield, Pascale, Piacentini, Pogosyan,
  Prunet, Rao, Romeo, Ruhl, Scaramuzzi, Sforna, \& Vittorio}]{melnat2000}
de~Bernardis, P., Ade, P.~A.~R., Bock, J.~J., {et~al.} 2000, \nat, 404, 955

\bibitem[{Douspis {et~al.}(2003)Douspis, Riazuelo, Zolnierowski, \&
  Blanchard}]{nous2003}
Douspis, M., Riazuelo, A., Zolnierowski, Y., \& Blanchard, A. 2003, \aap, 405,
  409

\bibitem[{Einstein(1917)}]{eins17}
Einstein, A. 1917, Preuss.\ Akad.\ Wiss.\ Sitz.\, 142

\bibitem[{Eisenstein {et~al.}(2005)Eisenstein, Zehavi, Hogg, Scoccimarro,
  Blanton, Nichol, Scranton, Seo, Tegmark, Zheng, Anderson, Annis, Bahcall,
  Brinkmann, Burles, Castander, Connolly, Csabai, Doi, Fukugita, Frieman,
  Glazebrook, Gunn, Hendry, Hennessy, Ivezic, Kent, Knapp, Lin, Loh, Lupton,
  Margon, McKay, Meiksin, Munn, Pope, Richmond, Schlegel, Schneider, Shimasaku,
  Stoughton, Strauss, SubbaRao, Szalay, Szapudi, Tucker, Yanny, \&
  York}]{eisenstein2005}
Eisenstein, D.~J., Zehavi, I., Hogg, D.~W., {et~al.} 2005, \apj, 633, 560

\bibitem[{{Elgar{\o}y} \& {Multam{\"a}ki}(2007)}]{elga07}
{Elgar{\o}y}, O. \& {Multam{\"a}ki}, T. 2007, \aap, 471, 65

\bibitem[{Elizalde {et~al.}(2004)Elizalde, Nojiri, \& Odintsov}]{st1}
Elizalde, E., Nojiri, S., \& Odintsov, S.~D. 2004, \prd, 70, 043539

\bibitem[{Fosalba {et~al.}(2003)Fosalba, Gaztanaga, \& Castander}]{corrisw}
Fosalba, P., Gaztanaga, E., \& Castander, F. 2003, \apj, 597, L89

\bibitem[{{Hunt} \& {Sarkar}(2007)}]{HuntSarkar07}
{Hunt}, P. \& {Sarkar}, S. 2007, ArXiv e-prints, 706

\bibitem[{{Kim} {et~al.}(2004){Kim}, {Linder}, {Miquel}, \& {Mostek}}]{kim04}
{Kim}, A.~G., {Linder}, E.~V., {Miquel}, R., \& {Mostek}, N. 2004, \mnras, 347,
  909

\bibitem[{{Lewis} \& {Bridle}(2002)}]{cosmomc}
{Lewis}, A. \& {Bridle}, S. 2002, \prd, 66, 103511

\bibitem[{Lewis {et~al.}(2000)Lewis, Challinor, \& Lasenby}]{CAMB}
Lewis, A., Challinor, A., \& Lasenby, A. 2000, \apj, 538, 473, {\rm CAMB} code
  freely available at http URL http://camb.info/

\bibitem[{Linder \& Huterer(2005)}]{pca4}
Linder, E.~V. \& Huterer, D. 2005, \prd, 72, 043509

\bibitem[{Lineweaver {et~al.}(1997)Lineweaver, Barbosa, Blanchard, \&
  Bartlett}]{Lineweaver1997}
Lineweaver, C.~H., Barbosa, D., Blanchard, A., \& Bartlett, J.~G. 1997, \aap,
  322, 365

\bibitem[{Melchiorri {et~al.}(2003)Melchiorri, Mersini, {\"O}dman, \&
  Trodden}]{melch}
Melchiorri, A., Mersini, L., {\"O}dman, C.~J., \& Trodden, M. 2003, \prd, 68,
  043509

\bibitem[{{\"O}dman {et~al.}(2004){\"O}dman, Hobson, Lasenby, \&
  Melchiorri}]{odman2004}
{\"O}dman, C., Hobson, M., Lasenby, A., \& Melchiorri, A. 2004, Int.\ J.\ Mod.\
  Phys.\, D13, 1661

\bibitem[{Peebles \& Ratra(2003)}]{quintinf}
Peebles, P.~J.~E. \& Ratra, B. 2003, Rev.\ Mod.\ Phys.\, 75, 559

\bibitem[{Perlmutter {et~al.}(1999)Perlmutter, Aldering, Goldhaber, Knop,
  Nugent, Castro, Deustua, Fabbro, Goobar, Groom, Hook, Kim, Kim, Lee, Nunes,
  Pain, Pennypacker, Quimby, Lidman, Ellis, Irwin, McMahon, Ruiz-Lapuente,
  Walton, Schaefer, Boyle, Filippenko, Matheson, Fruchter, Panagia, Newberg, \&
  Couch}]{perl98}
Perlmutter, S., Aldering, G., Goldhaber, G., {et~al.} 1999, \apj, 517, 565

\bibitem[{Pogosian(2005)}]{corrisw3}
Pogosian, L. 2005, JCAP, 0504, 015

\bibitem[{Ratra \& Peebles(1988)}]{quint888}
Ratra, B.~V. \& Peebles, P.~J.~E. 1988, \prd, 37, 3406

\bibitem[{{Riazuelo} \& {Uzan}(2002)}]{DEpert}
{Riazuelo}, A. \& {Uzan}, J.-P. 2002, \prd, 66, 023525

\bibitem[{{Riess} {et~al.}(1998){Riess}, {Filippenko}, {Challis},
  {Clocchiatti}, {Diercks}, {Garnavich}, {Gilliland}, {Hogan}, {Jha},
  {Kirshner}, {Leibundgut}, {Phillips}, {Reiss}, {Schmidt}, {Schommer},
  {Smith}, {Spyromilio}, {Stubbs}, {Suntzeff}, \& {Tonry}}]{riess98}
{Riess}, A.~G., {Filippenko}, A.~V., {Challis}, P., {et~al.} 1998, \aj, 116,
  1009

\bibitem[{{Riess} {et~al.}(2007){Riess}, {Strolger}, {Casertano}, {Ferguson},
  {Mobasher}, {Gold}, {Challis}, {Filippenko}, {Jha}, {Li}, {Tonry}, {Foley},
  {Kirshner}, {Dickinson}, {MacDonald}, {Eisenstein}, {Livio}, {Younger}, {Xu},
  {Dahl{\'e}n}, \& {Stern}}]{Riess07}
{Riess}, A.~G., {Strolger}, L.-G., {Casertano}, S., {et~al.} 2007, \apj, 659,
  98

\bibitem[{{Spergel} {et~al.}(2007){Spergel}, {Bean}, {Dor{\'e}}, {Nolta},
  {Bennett}, {Dunkley}, {Hinshaw}, {Jarosik}, {Komatsu}, {Page}, {Peiris},
  {Verde}, {Halpern}, {Hill}, {Kogut}, {Limon}, {Meyer}, {Odegard}, {Tucker},
  {Weiland}, {Wollack}, \& {Wright}}]{wmap2par}
{Spergel}, D.~N., {Bean}, R., {Dor{\'e}}, O., {et~al.} 2007, \apjs, 170, 377

\bibitem[{Spergel {et~al.}(2003)Spergel, Verde, Peiris, Komatsu, Nolta,
  Bennett, Halpern, Hinshaw, Jarosik, Kogut, Limon, Meyer, Page, Tucker,
  Weiland, Wollack, \& Wright}]{spergel03}
Spergel, D.~N., Verde, L., Peiris, H.~V., {et~al.} 2003, \apjs, 148, 175

\bibitem[{Wetterich(1988)}]{quint88}
Wetterich, C. 1988, \aap, 301, 321

\bibitem[{{Wood-Vasey} {et~al.}(2007){Wood-Vasey}, {Miknaitis}, {Stubbs},
  {Jha}, {Riess}, {Garnavich}, {Kirshner}, {Aguilera}, {Becker}, {Blackman},
  {Blondin}, {Challis}, {Clocchiatti}, {Conley}, {Covarrubias}, {Davis},
  {Filippenko}, {Foley}, {Garg}, {Hicken}, {Krisciunas}, {Leibundgut}, {Li},
  {Matheson}, {Miceli}, {Narayan}, {Pignata}, {Prieto}, {Rest}, {Salvo},
  {Schmidt}, {Smith}, {Sollerman}, {Spyromilio}, {Tonry}, {Suntzeff}, \&
  {Zenteno}}]{Wood07}
{Wood-Vasey}, W.~M., {Miknaitis}, G., {Stubbs}, C.~W., {et~al.} 2007, \apj,
  666, 694

\end{thebibliography}

\end{document}